\documentclass[preprint,aps,tightenlines,floatfix,showpacs]{revtex4}
\usepackage{graphics}
\usepackage{bm}
\usepackage{amsmath}
\usepackage{amssymb}

\begin{document}
  \begin{center}
    {\large\bf Elastic Electron Scattering off Exotic Nuclei} \\ \vskip 5 mm
    C.A. Bertulani$^{(a)}$ \\
    $^{(a)}$ Department of Physics, University of Arizona, Tucson, AZ 85721 
  \end{center}

\section{Introduction}
The use of radioactive nuclear beams produced by fragmentation in high-energy
heavy-ion reactions lead to the discovery of halo nuclei, such as $^{11}$Li,
about 20 years ago \cite{Ta85}. Nowadays a huge number of $\beta$-unstable
nuclei far from stability are being studied thanks to further technical
improvements. Unstable nuclei far from stability are known to play an
important role in nucleosynthesis. Detailed studies of the structure and their
reactions will have unprecedented impact on astrophysics \cite{BHM02}.

The first experiments with unstable nuclear beams aimed at determining nuclear
sizes by measuring the interaction cross section in high energy collisions
\cite{Ta85}. Successive use of this technique has yielded nuclear size data
over a wide range of isotopes. Other techniques, e.g. isotope-shift
measurements, have allowed to extract the charge size. The growth of a neutron
skin with the neutron number in several isotopes have been deduced from
nuclear- and charge-size data \cite{OST01}.

Undoubtedly,  electron scattering off nuclei would provide the
most direct information about charge distribution, which is closely related to
the spatial distribution of protons \cite{Hof61}.
A technical proposal for an electron-heavy-ion collider has been incorporated
in the GSI/Germany physics program \cite{Haik05}. A similar program exists for
the RIKEN/Japan facility \cite{Sud01}. In both cases the main purpose is to
study the structure of nuclei far from the stability line. The advantages of
using electrons in the investigation of the nuclear structure are mainly
related to the fact that the electron-nucleus interaction is relatively weak.
For this reason multiple scattering effects are usually neglected and the
scattering process is described in terms of perturbation theory. Since the
reaction mechanism in perturbation theory is well under control the connection
between the cross section and quantities such as charge distributions,
transition densities, response functions etc., is well understood \cite{Hof56}.

Under the impulse approximation, or plane wave Born approximation, the charge
form factor can be determined from the differential cross section of elastic
electron scattering. Since the charge distribution, $\rho_{ch}(r)$, is
obtained from the charge form factor by a Fourier transformation, one can
experimentally determine $\rho_{ch}(r)$ by differential cross-section
measurements covering a wide range of momentum transfer $q$. One can obtain
information on the size and diffuseness when the charge form factor is
measured at least up to the first maximum. To do this within a reasonable
measuring time of one week, a luminosity larger than 10$^{26}$ cm$^{-2}%
$s$^{-1}$ is required, for example for the $^{132}$Sn isotope \cite{Haik05}.

On the theoretical side the difference between the proton and neutron
distributions can be obtained in the framework of Hartree-Fock (HF) method
(see for example \cite{HL98}) or Hartree-Fock-Bogoliubov (HFB) method (see for
example \cite{Miz00,Ant05}). As a rule of thumb, a theoretical calculation of
the nuclear density is considered good when it reproduces the data on elastic
electron scattering. But some details of the theoretical densities might not
be accessible in the experiments, due to poor resolution or limited
experimental reach of the momentum transfer $q$.

\section{Elastic Electron Scattering}

In the plane wave Born approximation (PWBA) the relation between the charge
density and the cross section is given by%
\begin{equation}
\left(  \frac{d\sigma}{d\Omega}\right)  _{\mathrm{PWBA}}=\frac{\sigma_{M}%
}{1+\left(  2E/M_{A}\right)  \sin^{2}\left(  \theta/2\right)  }\ |F_{ch}%
\left(  q\right)  |^{2}, \label{PWBA}%
\end{equation}
where $\sigma_{M}=(Z^{2}\alpha^{2}/4E^{2})\cos^{2}\left(  \theta/2\right)
\sin^{-4}\left(  \theta/2\right)  $ is the Mott cross section, the term in the
denominator is a recoil correction, $E$ is the electron total energy, $M_{A}$
is the mass of the nucleus and $\theta$\ is the scattering angle.

The charge form factor $F_{ch}\left(  q\right)  $ for a spherical mass
distribution is given by%
\begin{equation}
F_{ch}\left(  q\right)  =4\pi\int_{0}^{\infty}dr\ r^{2}j_{0}\left(  qr\right)
\rho_{ch}\left(  r\right)  , \label{form}%
\end{equation}
where $q=2k\sin\left(  \theta/2\right)  $ is the momentum transfer, $\hbar k$
is the electron momentum, and $E=\sqrt{\hbar^{2}k^{2}c^{2}+m_{e}^{2}c^{4}}$.
The low momentum expansion of eq. \ref{form} yields the leading terms
$F_{ch}\left(  q\right)  /Z=1-{q^{2}}\left\langle r_{ch}^{2}%
\right\rangle/6 +\cdots. 
$
Thus, a measurement at low momentum transfer yields a direct assessment of the
mean square radius of the charge distribution, $\left\langle r_{ch}%
^{2}\right\rangle ^{1/2}$. However, as more details of the charge distribution
is probed more terms of this series are needed and, for a precise description
of it, the form factor dependence for large momenta $q$ is needed.

A theoretical calculation of the charge density entering eq. \ref{form} can be
obtained in many ways. Let $\rho_{p}\left(  \mathbf{r}\right)  $ and $\rho
_{n}\left(  \mathbf{r}\right)  $\ denote the point distributions of the
protons and the neutrons, respectively, as calculated, e.g. from
single-particle wavefunctions obtained from an average one-body potential
well, the latter in general being different for protons and neutrons. If
$f_{Ep}\left(  \mathbf{r}\right)  $ and $f_{En}\left(  \mathbf{r}\right)  $
are the spatial charge distributions of the proton and the neutron in the
non-relativistic limit, then the charge distribution of the nucleus is given
by%
\begin{equation}
\rho_{ch}\left(  \mathbf{r}\right)  =\int\rho_{p}\left(  \mathbf{r}^{\prime
}\right)  \ f_{Ep}\left(  \mathbf{r-r}^{\prime}\right)  \ d^{3}r^{\prime}%
+\int\rho_{n}\left(  \mathbf{r}^{\prime}\right)  \ f_{En}\left(
\mathbf{r-r}^{\prime}\right)  \ d^{3}r^{\prime}. \label{rhoch}%
\end{equation}

The second term on the right-hand side of eq. \ref{rhoch} plays an important
role in the interpretation of the charge distribution of some nuclear
isotopes. For example, the half-density charge radius increases 2\% from
$^{40}$Ca to $^{48}$Ca, whereas the surface thickness decreases by 10\% with
the result that there is more charge in the surface region of $^{40}$Ca than
of $^{48}$Ca \cite{Fro68}. This also results that the rms charge radius of
$^{48}$Ca is slightly smaller than that of $^{40}$Ca. The reason for this
anomaly is that the added $f_{7/2}$ neutrons contribute negatively to the
charge distribution in the surface and more than compensate for the increase
in the rms radius of the proton distribution.

For the proton the charge density $f_{Ep}\left(  r\right)  $ in eq.
\ref{rhoch} is taken as an exponential function, what corresponds to a form
factor $G_{Ep}(q)\equiv$ $\widetilde{f}_{Ep}\left(  q\right)  =(1+q^{2}%
/\Lambda^{2})^{-1}$. For the neutron a good parametrization is $G_{En}%
(q)=-\mu_{n}\tau G_{Ep}(q)/(1+p\tau)$, where $\mu_{n}$ is the neutron magnetic
dipole moment and $\tau=q^{2}/4m_{N}$. One can use $\Lambda^{2}=0.71 $
fm$^{-2}$ (corresponding to a proton rms radius of 0.87 fm) and $p=5.6$, which
Galster et al. \cite{Gal71} have shown to reproduce electron-nucleon
scattering data.

Eqs. \ref{PWBA}-\ref{rhoch} are based on the first Born
approximation. They give good results for light nuclei (e.g.
$^{12}$C) and high-energy electrons. For large-$Z$ nuclei the
agreement with experiments is only of a qualitative nature. The
effects of distortion of the electron waves have been studied by
many authors (see, e.g. ref. \cite{MF48,Yenn53,Cut67}). More
important than the change in the normalization of the cross section
is the displacement of the minima. It is well known that a very
simple modification can be done in the PWBA equation that reproduces
the shift of the minima to lower $q$'s. One replaces the momentum
transfer $q$ in the form factor of eq. \ref{PWBA} with the effective
momentum transfer $q_{eff}=q\left(  1+3Ze^{2}/2R_{ch}E\right) $,
where $E$ is the electron energy and $R\simeq1.2\ A^{-1/3}$ fm. This
is because a measurement at momentum transfer $q$ probes in fact \
$|F(q)|^{2}$ at $q=q_{eff}$ due to the attraction the electrons feel
by the positive charge of the nucleus. This expression for $q_{eff}$
assumes a homogeneous distribution of charge within a sphere of
radius $R$.

A realistic description of the elastic electron scattering cross section
requires the full solution of the Dirac equation. The Dirac equation for the
elastic scattering from a charge distribution can be found in standard
textbooks, e.g. \cite{EG88}. Numerous DWBA codes based on Dirac distorted
waves have been developed and are public. 

\section{Skins and Halos}

\subsection{Neutron Skins}

Appreciable differences between neutron and proton radii are
expected \cite{Dob96} to characterize the nuclei at the border of
the stability line. The liquid drop formula expresses the binding
energy of a nucleus with $N$ neutrons and $Z$ protons as a sum of
bulk,
surface, symmetry and Coulomb energies ${E/A}=-a_{V}A+a_{S}A^{2/3}%
+S(N-Z)^{2}/A+a_{C}Z^{2}/A^{1/3}\pm a_{p}A^{-1/2},$where $a_{V}$, $a_{S}$,
$a_{p}$, $S$ and $a_{C}$ are parameters fitted to the experimental data of
binding energy of nuclei. This equation does not distinguish between surface
($S$) and volume ($V$) symmetry energies. As shown in ref. \cite{Da03}, this
can be achieved by partitioning the particle asymmetry as $N-Z=N_{S}%
-Z_{S}+N_{V}-Z_{V}$. The total symmetry energy $S$ \ then takes on
the form
$S=S_{V}(N_{V}-Z_{V})^{2}/A+S_{S}(N_{S}-Z_{S})^{2}/A^{2/3}.$
Minimizing under fixed $N-Z$ leads to an improved liquid drop
formula \cite{Da03} with the term $S_{V}(N-Z)^{2}/A$ replaced by
$
S_{V}(N-Z)^{2}/A[1+\left(  S_{V}/S_{S}\right)  A^{-1/3}]$.
The same approach also yields a relation between the neutron skin $
R_{np}=R_{n}-R_{p}$, and $S_{S}$, $S_{V}$, namely \cite{Da03}%
${R_{n}-R_{p}}/{R}=({A}/{6NZ})
[{N-Z-\left(  a_{C}/12S_{V}\right)  ZA^{2/3}}]/[{1+\left(  S_{S}/S_{V}\right)
A^{1/3}}], $
where $R=(R_{n}+R_{p})/2$.

Here the Coulomb contribution is essential; e.g. for $N=Z$ \ the neutron skin
$R_{np}$ is negative due to the Coulomb repulsion of the protons. A wide
variation of values of $S_{V}$ and $S_{S}$ can be found in the literature.
These values have been obtained by comparing the above predictions for energy
and neutron skin to theoretical calculations of nuclear densities and
experimental data on other observables \cite{Da03,St05}. For heavy nuclei,
with $A\gg1$, $NZ\simeq A^{2}/2$, and using $a_{C}=0.69$ MeV, $S_{V}=28$ MeV
and $R=1.2A^{1/3}$ fm, one gets%
\begin{equation}
R_{np}=R_{n}-R_{p}\simeq0.4\left(  \frac{S_{V}}{S_{S}}\right)  \left(
\delta-2.05\times10^{-3}ZA^{-1/3}\right)  \text{ fm},\label{Pawelh}%
\end{equation}
where $\delta=\left(  N-Z\right)  /A$ is the asymmetry parameter. If
one assumes that the central densities for neutrons and protons are
roughly the same and that they are both described by a uniform
distribution with sharp-cutoff radii, $R_{n}$ and $R_{p}$, one finds
$R_{np}\simeq 0.8A^{1/3}\delta$ fm. Of course, the sharp sphere
model is too simple.

On the experimental front, a study of antiprotonic atoms published
in reference \cite{Trz01} obtained the following fitted formula for
the neutron skin of stable nuclei in terms of the root mean square
(rms) radii of protons and neutrons
\begin{equation}
\Delta r_{np}=\left\langle r_{n}^{2}\right\rangle ^{1/2}-\left\langle
r_{p}^{2}\right\rangle ^{1/2}=\left(  -0.04\pm0.03\right)  +\left(
1.01\pm0.15\right)  \delta\ \text{fm.}\label{delta}%
\end{equation}
The relation of the mean square radii with the half-density radii is given by
$\left\langle r_{n}^{2}\right\rangle =3R_{n}^{2}/5+7\pi^{2}a_{n}^{2}/5$, where
$a_{n}$ is the diffuseness parameter. For heavy nuclei, assuming $a_{n}%
=a_{p}\ll R_{n},R_{p}$, one gets the same linear dependence on the asymmetry
parameter as in eq. \ref{Pawelh}.

Eq. \ref{delta} can be used as the starting point for
accessing the dependence of electron scattering on the neutron skin
of heavy nuclei. Applying it to calcium isotopes as an example, one
obtains that the neutron skin varies from $-0.15$ fm for $^{35}$Ca
(proton-rich with negative neutron skin) to $0.25$ fm for $^{53}$Ca.
A negative neutron skin obviously means an excess of protons at the
surface. 

For heavy nuclei the charge and neutron distributions can be described by a
Fermi distribution. The diffuseness is usually much smaller than the
half-density radius, $a_{p,n}\ll R_{p,n}$. The neutron skin is then given by
$R_{np}=R_{n}-R_{p}\simeq\sqrt{5/3}\ \Delta r_{np}$. One can assume further that
the nuclear charge radius is given by $R_{p}=1.2A^{1/3}%
\ \mathrm{fm}-R_{np}/2$ with $\Delta r_{np}$ given by eq. \ref{delta}.  
The first and second minima of the form
factors occur at $q_{1}=4.49/R_{p}$ and $q_{2}=7.73/R_{p}$, respectively,
corresponding to the zeroes of the transcendental equation $\tan\left(
qR_{p}\right)  =qR_{p}$.

The linear dependence of $R_{p}$ with the neutron skin (and with the asymmetry
parameter $\delta$), also imply a linear dependence of the position of the
minima,
\begin{equation}
q_{1}\simeq\frac{3.74}{A^{1/3}}\ \ \left[  1-0.535\frac{(N-Z)}{A^{4/3}%
}\right]  ^{-1}\ \mathrm{fm}^{-1},\ \ \ \ \ \ \ \ \mathrm{and}%
\ \ \ \ \ \ q_{2}=1.72\ q_{1}\ .
\end{equation}
For $^{100}$Sn the first minimum is expected to occur at $q_{1}=0.806$
$\mathrm{fm}^{-1}=159$ MeV/c, while for $^{132}$Sn it occurs at $q_{1}=0.754$
$\mathrm{fm}^{-1}=149$ MeV/c.

The variation of
$q_{1}$ with the neutron skin of neighboring isotopes, $\Delta q_{1}%
\simeq2/A^{8/3}$ fm$^{-1}$,  is too small to be measured accurately. The first
minimum, $q_{1}$, changes from $220$ MeV/c for $^{35}$Ca to $204$ MeV/c for
$^{53}$Ca, an approximate 7\%, which is certainly within the experimental
resolution. Of course, sudden changes of the neutron skin with $\delta$\ might
happen due to shell closures, pairing,  and other microscopic effects.

To be more specific, let us assume that a reasonable goal is to
obtain accurate results for the charge radius $\left\langle
r_{p}^{2}\right\rangle ^{1/2},$ so that $\delta\left\langle
r_{p}^{2}\right\rangle ^{1/2}<0.05$ fm. This implies that the
measurement of $q_{1}$ has to be such that $\left( \Delta
q_{1}/q_{1}\right)  <q_{1}\left[  \text{fm}^{-1}\right]  \%$, with
$q_{1}$ in units of fm$^{-1}$ and the right-hand side of the
inequality yielding the percent value. For $^{53}$Ca, q$_{1}=$
$1.11$ fm$^{-1}$ meaning that the experimental resolution on the
value of $q_{1}$ has to be within 1\% if $\delta\left\langle
r_{p}^{2}\right\rangle ^{1/2}<0.05$ fm is a required precision. 
Of course,
the ultimate test of a given theoretical model will be a good
reproduction of the measured data, below and beyond the first
minimum.

\subsection{Neutron halos}

Elastic electron scattering will be very important to determine
charge distributions in proton-rich nuclei. This will complement the
basic information on the charge distribution in, e.g. $^{8}$B,
obtained in nucleon knockout reactions \cite{Schw95}. For light
nuclei composed by a core nucleus and an extended distribution of
halo nucleons, the nuclear matter form factor can be fitted with the
simple
expression $
F(q)=\left(  1-g\right)  \exp\left(  -q^{2}a_{1}^{2}/4\right)  +
{g}/({1+a_{2}^{2}q^{2}}),$
with the density normalized to one. $g$ is the fraction of nucleons
in the halo. In this expression the first term \ follows from the
assumption that the core is described by a Gaussian and the halo
nucleons by an Yukawa distribution. Taking $^{11}$Li as an example,
the following set of parameters can be used $g=0.18,$ $a_{1}=2.0$ fm
and $a_{2}=6.5$ fm. Although only few nucleons are in the halo they
change dramatically the appearance of the squared form factor. 
Even
when the individual contribution of the halo nucleons is small and
barely visible in a linear plot of the matter distribution, it is
very important for the form factor of the total matter distribution.
It is responsible for the narrow peak which develops at low momentum
transfers. This signature of the halo was indeed the motivation for
the early experiments with radioactive beams. The narrow peak was
observed in momentum distributions following knockout reactions
\cite{Ta85}.

Elastic electron scattering will not be sensitive to the narrow peak of
$\left\vert F(q)\right\vert ^{2}$ at small momentum $q$, but to the form
factor of the charge distribution, $\left\vert F_{ch}(q)\right\vert ^{2}$. 
The determination of this form factor will tell us if the core
has been appreciably modified due to the presence of the halo nucleons.

In order to explain the spin, parities, separation energies and size
of exotic nuclei consistently a microscopic calculation is needed.
One possibility is to resort to a Hartree-Fock (HF) calculation.
Unfortunately, the HF theory cannot provide the predictions for the
separation energies within the required accuracy of hundred keV.
I have used a simple and tractable HF\ method ~\cite{BBS89} to
generate synthetic data for the charge-distribution of $^{11}$Li.
Details of this method is described in ref. \cite{SB97}. Assuming
spherical symmetry, the equation for the Skyrme interaction can be
written as \
\begin{equation}
\left[  -\mathbf{\nabla}\frac{\hbar^{2}}{2m^{\ast}(r)}\mathbf{\nabla
}+V(\mathbf{r})\right]  \psi_{i}(\mathbf{r})=\epsilon_{i}\psi_{i}%
(\mathbf{r})\label{schro}%
\end{equation}
where $\,m^{\ast}(r)\,$ is the effective mass. The potential $\,V(\mathbf{r}%
)\,$ has a central, a spin-orbit and a Coulomb term.
The central potential is multiplied by a constant factor $\,f\,$ only for the
last neutron configuration:%
\begin{equation}
V_{\mathrm{central}}(r)=fV_{HF}(r),\left\{
\begin{array}
[c]{l}%
f\neq1\quad\mathrm{for\;last\;neutron\;configuration}\\
f=1\quad\mathrm{otherwise.}%
\end{array}
\right.  \label{vcentral}%
\end{equation}

As the effective interaction, a parameter set of the density dependent Skyrme
force, so called BKN interaction \cite{BKN}, is adopted. The parameter set of
BKN interaction has the effective mass m$^{\ast}$/m =1 and gives realistic
single particle energies near the Fermi surface in light nuclei. The original
BKN force has no spin-orbit interaction. In the calculations, the
spin-orbit term was introduced 
in the interaction so that the single-particle energy of the
last neutron orbit becomes close to the experimental separation energy. In
this way, the asymptotic form of the loosely-bound wave function becomes
realistic in the neutron-rich nucleus. In order to obtain the nuclear sizes,
the \ rms \ radii of the occupied nucleon orbits are multiplied by the shell
model occupation probabilities, which are also obtained in the calculations.

The elastic form factor for the matter distribution obtained in the HF
calculations are very close to the ones calculated by the empirical formula
given at the beginning of this section.  The lack of minima, and of secondary peaks (as in the empirical formula), make it difficult to extract from $\left\vert F_{ch}%
(q)\right\vert ^{2}$ more detailed information on the charge-density
profile. For example, in the case of $^{6}$Li a good fit to
experimental data was obtained with \cite{Sue67}
$
\left\vert F_{ch}(q)\right\vert ^{2}\propto\exp(-a^{2}q^{2})-Cq^2\exp(-b^{2}%
q^{2}),
$
with $a=0.933$ fm, $b=1.3$ fm, and $C=0.205$. However, the data cannot be
fitted by using a model in which the nucleons move in a single-particle potential.

\subsection{Proton Halos}

I will consider  $^{8}$B as a prototype of proton halo nucleus.
This nucleus is perhaps the most likely candidate for having a
proton halo structure, as its last proton has a binding energy of
only 137 keV. The charge density for this nucleus can be calculated
in the framework of the Skyrme HF model. I use here the
results obtained in ref. \cite{Shyam}, where axially symmetric HF
equations were used with SLy4 \cite{Chab97} Skyrme interaction which
has been constructed by fitting the experimental data on radii and
binding energies of symmetric and neutron-rich nuclei. Pairing
correlations among nucleons have been treated within the BCS pairing
method. The form factor squared for the charge density in $^{8}$B is
calculated.

The width of the charge form factor squared  
is $\Delta_{ch}=0.505$ fm$^{-1}=99.6$ MeV/c. The
corresponding values for the neutron and the total matter distributions are,
respectively, $\Delta_{n}=0.512$ fm$^{-1}=101$ MeV/c and $\Delta_{tot}=0.545$
fm$^{-1}=108$ MeV/c. This amounts to approximately 10\% differences of matter
and charge form factors in $^{8}$B. 

The proton halo in $^{8}$B is mainly due to the unpaired proton in
the p$_{3/2}$ orbit. It is clear that for such a narrow halo the
size of the nucleon also matters. A slice of the nucleon is
included in a thin spherical shell of radius $r$ and thickness $dr$
from the center of the nucleus. If the position of the nucleon is
given by $R$, the part of the proton charge included in the
spherical shell is given by
\begin{equation}
d\rho_{ch}=2\pi r^{2}dr\int_{0}^{\pi}d\theta\ \rho_{p}\left(  \mathbf{x}%
\right)  \sin\theta,\label{drhon}%
\end{equation}
where $\rho_{p}\left(  \mathbf{x}\right)  $ is the charge distribution inside
a proton at a distance $\mathbf{x}$ from its center. The coordinates  
are related by $x^{2}=r^{2}+R^{2}-2rR\cos\theta
$. The contribution to the nuclear charge distribution from a single-proton in
this spherical shell is thus given by $\mathcal{N}_{p}\left(  R,r\right)
=d\rho_{ch}/4\pi r^{2}dr.$

Assuming that the charge distribution of the proton is described either by a
Gaussian or a Yukawa form, the integral in eq. \ref{drhon} can be performed
analytically, yielding%
\begin{equation}
\mathcal{N}_{p}^{(G)}\left(  R,r\right)  =\frac{1}{4\pi^{1/2}arR}\left\{
\begin{array}
[c]{c}%
\frac{1}{\pi}\left\{  \exp\left[  -\frac{\left(  R-r\right)  ^{2}}{a^{2}%
}\right]  -\exp\left[  -\frac{\left(  R+r\right)  ^{2}}{a^{2}}\right]
\right\}  ,\ \text{\ \ \ for a Gaussian dist.}\\
\frac{1}{2}\left\{  \exp\left[  -\frac{\left\vert R-r\right\vert }{a}\right]
-\exp\left[  -\frac{\left\vert R+r\right\vert }{a}\right]  \right\}
,\ \text{\ \ \ for a Yukawa dist.,}%
\end{array}
\right.  \label{npc}%
\end{equation}
where $a$ is the proton radius parameter.

The charge distribution at the surface of a heavy proton-rich nucleus,
$\delta\rho_{ch}\left(  r\right)  $, may be described as a pile-up of protons
forming a skin. Let $n_{i}$ be the number of protons in the skin and $R_{i}$
their distance to the center of the nucleus. One gets%
\begin{equation}
\delta\rho_{ch}\left(  r\right)  =\sum_{i}n_{i}\mathcal{N}_{p}\left(
r,R_{i}\right)  .\label{npcs}%
\end{equation}
Assuming $R_{i}$ to be constant, equal to the nuclear charge radius $R$, and
using eq. \ref{npc} it is evident that while the density at the surface
increases, its size $R$ and width $a$, remain unaltered. \ The form factor
associated with this charge distribution is given by%
\begin{equation}
\delta F_{ch}\left(  q\right)  =\frac{4\pi}{q}\sum_{i}n_{i}\int_{0}^{\infty
}dr\ r\ \mathcal{N}_{p}\left(  r,R_{i}\right)  \ \sin\left(  qr\right)
=\exp\left(  -qa^{2}\right)  \sum_{i}n_{i}\frac{\sin\left(  qR_{i}\right)
}{qR_{i}},\label{formg}%
\end{equation}
where the last result is for the Gaussian distribution. An analytical
expression can also be obtained for the Yukawa distribution. For $R_{i}=R$,
expression \ref{formg} shows that the increase of density in the skin does not
change the shape of the form factor, or of the cross section, but just its
normalization. The fall down of the cross section is determined by $a$ alone,
and not by $n_{i}$. \ If the charge of additional protons is distributed
homogeneously across the nucleus including the skin, the normalized form
factor will not change, except for a small change in $R$.

For a proton halo nucleus it is more appropriate to replace $\sum_{i}%
n_{i}\rightarrow4\pi\int dR\ R^{2}\ \mathcal{N}_{p}\left(  r,R\right)
\delta\rho_{ch}\left(  R\right)  $, where $\delta\rho_{ch}\left(  R\right)  $
is the density change created by the extended wavefunction of the halo
protons. One then recast eq. \ref{formg} in the form%
\begin{equation}
\delta F_{ch}\left(  q\right)  =\frac{4\pi}{q}\exp\left(  -qa^{2}\right)
\int_{0}^{\infty}dR\ R\ \delta\rho_{ch}\left(  R\right)  \ \sin\left(
qR\right)  .\label{formh}%
\end{equation}
The shape of the form factor is here dependent not only on the
proton size but also on the details of the halo density
distribution. For $^{8}$B, the halo size is determined by the
valence proton in a p$_{3/2}$ orbit. The density
$\delta\rho_{ch}\left(  R\right)  $\ due this proton can be
calculated with a Woods-Saxon model. Using the same potential
parameters as in ref. \cite{NBC06} I compare the
form factor $\left\vert \delta F_{ch}\left( q\right)  \right\vert
^{2}$ to the charge form factor $\left\vert F_{ch}\left(
q\right)  \right\vert ^{2}$. The halo is found to contribute to 
a narrow form factor. However, in
contrast to the neutron halo case, the
charge form factor of $^{8}$B does not show a pronounced influence
of the halo charge distribution. 

The rms radius of the charge distribution can be calculated from 
$
\left\langle r_{ch}^{2}\right\rangle =-6\left[  {dF_{ch}}/{d\left(
q^{2}\right)  }\right]  _{q^{2}=0}. %
$
Applying this relationship to the $^8$B charge form factor 
we get $\left\langle r_{ch}^{2}\right\rangle
^{1/2}=2.82$ fm which is close to the experimental value
$\left\langle r_{ch}^{2}\right\rangle _{\exp}^{1/2}=2.76\pm0.08$ fm.
The shape of the charge form factor can also be described by a
Gaussian distribution with radius parameter $a=2.30$ fm. In contrast
to the case of $^{11}$Li, the proton halo
in $^{8}$B does not seem to build up a two-Gaussian shaped form
factor. This observation also seems to be compatible with the
momentum distributions of $^{7}$Be fragments in knockout reactions
using $^{8}$B projectiles in high energy collisions \cite{Schw95}.
Electron scattering experiments will help to further elucidate this
property of proton halos.

\section{Inverse scattering problem}

In PWBA the inverse scattering problem can be easily solved. It is possible to
extract the form factor from the cross section and then, with an inversion of
the Fourier transform, to get the charge density distribution%
\begin{equation}
\rho_{ch}\left(  r\right)  =\frac{1}{2\pi^{2}}\int_{0}^{\infty}dq\ q^{2}%
j_{0}\left(  qr\right)  F_{ch}\left(  q\right)  .\label{densinv}%
\end{equation}
The PWBA approximation can be justified
only for light nuclei (e.g. $^{12}$C) in the region far from the diffraction
zeros. For higher $Z$ values the agreement with experiment is only of a
qualitative nature.

It is very common in the literature to use a theoretical model for
$\rho _{ch}\left(  r\right)  $, e.g. the HF calculations discussed
in the previous sections and compare the calculated $F(q)$ with
experimental data. When the fit is \textquotedblleft
reasonable\textquotedblright\ (usually guided by the eye) the model
is considered a good one. However, whereas the theoretical
$\rho_{ch}\left(  r\right)  $ can contain useful information about
the central part of the density (e.g. bubble-like nuclei, with a
depressed central density), an excellent fit to the available
experimental data does not necessarily mean that the data is
sensitive to those details. The obvious reason is that short
distances are probed by larger values of $q$. Experimental data from
electron-ion colliders will suffer from limited accuracy at large
values of $q$, possibly beyond $q=1$ fm$^{-1}$. 

In order to obtain an unbiased \textquotedblleft
experimental\textquotedblright\ $\rho_{ch}\left(  r\right)  $ one usually
assumes that the density is expanded as
$
\rho_{ch}\left(  r\right)  =\sum_{n=1}^{\infty}a_{n}f_{n}\left(  r\right)
,\label{dens}%
$
where the basis functions $f_{n}(r)$ are drawn from any convenient complete
set and the expansion coefficients $a_{n}$ are adjusted to reproduce the
differential elastic cross section. The corresponding Fourier transform then
takes the form%
\begin{equation}
\widetilde{\rho}(q)\equiv F_{ch}\left(  q\right)  =\sum_{n=1}^{\infty}%
a_{n}\widetilde{f_{n}}\left(  q\right)  ,%
\ \ \ \ \ \widetilde{f_{n}}\left(  q\right)  =4\pi\int_{0}^{\infty}dr\ r^{2}j_{0}\left(
qr\right)  f_{n}\left(  r\right)  .\label{fnq}%
\end{equation}

Evidently the sum in $n$ has to be truncated and this produces an
error in the determination of the charge density distribution. Another problem
is that, as shown by eq. \ref{densinv}, the solution of the inverse scattering
problem requires an accurate determination of the cross section up to large
momentum transfers. Electron scattering experiments in electron-ion colliders
will be performed within a limited range of $q$ and this will produce an
uncertainty in the determination of the charge density distribution. 

Two bases have been found useful \cite{FN73} in the analysis
of electron or proton scattering data. The present discussion is limited to
spherical nuclei, but generalizations to deformed nuclei can be done. The
Fourier-Bessel (FB) expansion (i.e. with $f_{n}$ taken as spherical Bessel
functions) is useful because of the orthogonality relation between spherical
Bessel functions%
\begin{equation}
\int_{0}^{R_{\max}}dr\ r^{2}j_{l}\left(  q_{n}r\right)  j_{l}\left(
q_{m}r\right)  =\frac{1}{2}R_{c}^{3}\left[  j_{l+1}\left(  q_{n}R_{\max
}\right)  \right]  ^{2}\delta_{nm},\label{orthob}%
\end{equation}
where the $q_{n}$ are defined such as 
$
j_{l}\left(  q_{n}R_{\max}\right)  =0.\label{nodej}%
$
The FB basis implies that the charge density $\rho_{ch}(r)$ should be zero for
values of $r$ larger than $R_{\max}$. For example, the basis can be defined as
follows
\begin{equation}
f_{n}\left(  r\right)  =j_{0}\left(  q_{n}r\right)  \Theta\left(  R_{\max
}-r\right)  \text{, \ \ \ \ \ \ \ \ \ \ \ \ \ }\widetilde{f_{n}}\left(
q\right)  =\frac{4\pi\left(  -1\right)  ^{n}R_{\max}}{q^{2}-q_{n}^{2}}%
j_{0}\left(  qR_{\max}\right)  \text{, \ }\label{FB}%
\end{equation}
where $\Theta$ is the step function, \ $R_{\max}$ is the expansion radius and
$q_{n}=n\pi/R_{\max}$.

In principle it is possible to obtain the $a_{n}$ coefficients measuring
directly the cross section at the $q_{n}$ momentum transfer. If the form
factor (\ref{form}) is known at $q_{n}$, the coefficients $a_{n}$ can be
obtained inserting (\ref{orthob}) and (\ref{FB}) in the definition
(\ref{form}) of the form factor, leading to%
\begin{equation}
a_{n}=\frac{F_{ch}\left(  q_{n}\right)  }{2\pi R_{\max}^{3}\left[
j_{1}\left(  q_{n}R_{\max}\right)  \right]  ^{2}}. \label{anFB}%
\end{equation}

In general the cross sections \ are measured at $q$ values different from
$q_{n}$. Using the expansion (\ref{FB}) of the charge density one finds for
the form factor the relation%
\begin{equation}
F_{ch}\left(  q\right)  =\frac{4\pi}{q}\sum_{n}a_{n}\frac{\left(  -1\right)
^{n}}{q^{2}-q_{n}^{2}}\sin\left(  qR_{\max}\right)  .\label{formFB}%
\end{equation}
By fitting the experimental $F_{ch}(q)$ one obtains the $a_{n}$ parameters and
reconstruct the nuclear charge density. Not all $a_{n}$'s are needed. Since
the integral of the density, or $F\left(  q=0\right)  $, is fixed to the
charge number there is one less degree of freedom. Also, densities tend to
zero at large $r$. Thus another condition can be used, e.g. that the
derivative of the density is zero at $R_{\max}$. \ Thus, when one talks about
$n$ expansion coefficients one means in fact that only $n-2$ coefficients need
to be used in eq. \ref{formFB}.\ For experiments performed up to $q_{\max}$
the number of expansion coefficients needed for the fit is determined by
$n_{\max}\simeq q_{\max}R_{\max}/\pi$.

A disadvantage of the FB expansion is that a relatively large number of terms
is often needed to accurately represent a typical confined density, e.g. for
light nuclei. One can use other expansion functions which are invoke less
number of expansion parameters, e.g. the Laguerre-Gauss (LG) expansion,
\[
f_{n}(r)=e^{-\alpha^{2}}L_{n}^{1/2}\left(  2\alpha^{2}\right)
,\ \ \ \ \ \ \ \ \ \ \ \ \ \ \ \ \ \text{and}\ \ \ \ \widetilde{f_{n}}%
(q)=4\pi^{3/2}\beta^{3}\left(  -1\right)  ^{n}e^{-\gamma^{2}}L_{n}%
^{1/2}\left(  2\gamma^{2}\right)  ,
\]
where $\alpha=r/\beta$, $\gamma=q\beta/2$, and $L_{n}^{p}$ is the generalized
Laguerre polynomial. Another possibility is to use an expansion on Hermite (H)
polynomials. In both cases, the number of terms needed to provide a reasonable
approximation to the density can be minimized by choosing $\beta$ in
accordance with the natural radial scale. For light nuclei $\beta=1-2$ fm can
be chosen, consistent with the parametrization of their densities. Then the
magnitude of $a_{n}$ decreases rapidly with $n$, but the quality of the fit
and the shape of the density are actually independent of $\beta$ over a wide
range.

For real data, the expansion coefficients $a_{n}$ are obtained by minimizing%
\[
\chi^{2}=\sum_{i}\left(  \frac{y_{i}-y(q_{i},a_{n})}{\sigma_{i}}\right)
^{2},
\]
where $y(q_{i},a_{n})$ is the fitted value of the cross section (form factor)
with a set of coefficients $a_{n}$ and $y_{i}$ are the experimental data at
momentum $q_{i}$ with uncertainty $\sigma_{i}$.

Increasing the number of coefficients does not improve the
quality of the fit. It only
produces more oscillations of the density. The reason is that terms with
larger $n$'s are only needed to reproduce the data at larger values of
momentum transfer. The fit to the data for
$q<q_{\max}$ is not affected but the presence of these new terms introduces
oscillations in the charge distribution. A possible fix to this problem is to
include pseudodata in addition to experimental data. This method is well known
in the literature \cite{FN73}. The pseudodata are used to enforce constraints
and to estimate the incompleteness error associated with the limitation of
experimental data to a finite range of momentum transfer.

\vskip 0.5cm
The author is grateful to Haik Simon for beneficial discussions. This work was supported by the U.\thinspace S. Department of Energy under grant No. DE-FG02-04ER41338.

\end{document}